\documentclass[twocolumn,showpacs,preprintnumbers,amsmath,amssymb]{revtex4}

\usepackage{graphicx}
\usepackage{dcolumn}
\usepackage{bm}

\begin{document}


\title{Mechanism of Fermi Level Pinning at Metal/Germanium Interfaces}
\author{K. Kasahara,$^{1}$ S. Yamada,$^{1}$ K. Sawano,$^{2}$ M. Miyao,$^{1}$ and K. Hamaya$^{1,3}$\footnote{E-mail: hamaya@ed.kyushu-u.ac.jp} }
\affiliation{$^{1}$Department of Electronics, Kyushu University, 744 Motooka, Fukuoka 819-0395, Japan}
\affiliation{$^{2}$Advanced Research Laboratories, Tokyo City University, 8-15-1 Todoroki, Tokyo 158-0082, Japan}
\affiliation{$^{3}$PRESTO, Japan Science and Technology Agency, Sanbancho, Tokyo 102-0075, Japan}

\date{\today}

\begin{abstract}
The physical origin of Fermi level pinning (FLP) at metal/Ge interfaces has been argued over a long period. Using the Fe$_{3}$Si/Ge(111) heterostructure developed originally, we can explore electrical transport properties through atomically matched metal/Ge junctions. Unlike the conventional metal/$p$-Ge junctions reported so far, we clearly observe rectifying current-voltage characteristics with a measurable Schottky barrier height, depending on the contact area of the Fe$_{3}$Si/Ge(111) junction. These results indicate that one should distinguish between intrinsic and extrinsic mechanisms for discussing the formation of the Schottky barrier at metal/Ge interfaces. This study will be developed for understanding FLP for almost all the metal/semiconductor interfaces. 
\end{abstract}

\maketitle
The fundamental mechanism of the formation of the Schottky barrier at metal/semiconductor interfaces has so far been studied over a long period.\cite{Heine,Rowe,Spicer,Tung,Hasegawa,Tersoff,Heslinga,Tung2,Connelly} In particular, the controversy about the physical origin of Fermi level pinning (FLP) at the metal/semiconductor interfaces remains unsolved.\cite{Mohney} In order to interpret FLP, the metal-induced gap states (MIGS),\cite{Heine,Tersoff} caused by the decay of traveling wave function from the metal electrode into the band gap of the semiconductor, have been suggested. The MIGS are considered to be intrinsic mechanisms. In contrast, other extrinsic mechanisms such as defect-induced gap states at the interface have been discussed.\cite{Spicer,Hasegawa,Tung2} 

Development of high mobility germanium (Ge) based metal-oxide-semiconductor field effect transistors (MOSFETs) is required for breaking down the ultimate scaling limit of silicon (Si)-based conventional complementary metal-oxide-semiconductor transistors.\cite{Lee} However, FLP is one of the critical issues even at metal/Ge interfaces,\cite{Mohney} disturbing the development of high-performance MOSFETs with metallic source and drain contacts. In general, the influence of FLP on metal/Ge interfaces is stronger than that on metal/Si ones. The strong FLP always results in relatively high Schottky barrier height (SBH: $\Phi$$_{b} \sim$ 0.6 eV) for metal/$n$-Ge junctions and ohmic characteristics for metal/$p$-Ge ones, arising from the charge neutrality level close to the valence band edge of Ge.\cite{Dimoulas,Toriumi} 

Recently, effective alleviation of the strong FLP was also demonstrated by an insertion of ultra-thin insulating layers between a metal and Ge,\cite{Nishimura,Kobayashi} in which electrical properties of the metal/$n$-Ge and metal/$p$-Ge junctions were varied to ohmic and rectifying characteristics with increasing the thickness of the insertion layer. It seems that the observed thickness dependence is strongly related to the blocking of the penetration of the wave function of electrons in the band gap of Ge. Thus, one has believed that the strong FLP at metal/Ge interfaces is predominantly due to MIGS,\cite{Nishimura,Kobayashi,Wager} i.e., intrinsic mechanisms. On the other hand, several groups have discussed the influence of extrinsic mechanisms such as dangling bonds on the strong FLP.\cite{Lieten,Zhou2,Yamane,Thathachary,Lieten2} Therefore, exploring the mechanism of FLP at metal/Ge interfaces is not only important for demonstrating high-performance MOSFETs but also for understanding FLP for almost all the metal/semiconductor interfaces. 
 
In order to elucidate the origin of the strong FLP, we propose that the use of the high-quality Fe$_{3}$Si(111)/Ge(111) junction with an atomically matched interface is effective.\cite{Yamane} Fe$_{3}$Si has a Heusler-type crystal structure\cite{Moss,Niculescu} and there is almost no lattice mismatch between Fe$_{3}$Si (0.564 $\sim$ 0.566 nm) and Ge (0.565 nm). Furthermore, the atomic matching at the junction between Fe$_{3}$Si(111) and Ge(111) is theoretically perfect, as illustrated in Fig. 1 of Ref.\cite{Yamane}. To date, we clarified that, for Fe$_{3}$Si/$n$-Ge(111) junctions, the SBH is unexpectedly lower ($\Phi_\text{b}$ $\sim$ 0.46 eV) than that due to the strong FLP reported so far (0.55 $\sim$ 0.65 eV) and, for Fe$_{3}$Si/$p$-Ge(111) junctions, rectifying behavior in current-voltage ($I$-$V$) characteristics is seen below 170 K.\cite{Yamane} Though these results have implied the presence of an extrinsic mechanism of the strong FLP, we have not yet obtained its reliable evidence. At least, we should distinguish between intrinsic and extrinsic mechanisms for discussing the formation of the Schottky barrier at metal/Ge interfaces. 
\begin{figure}[t]
\includegraphics[width=8.5cm]{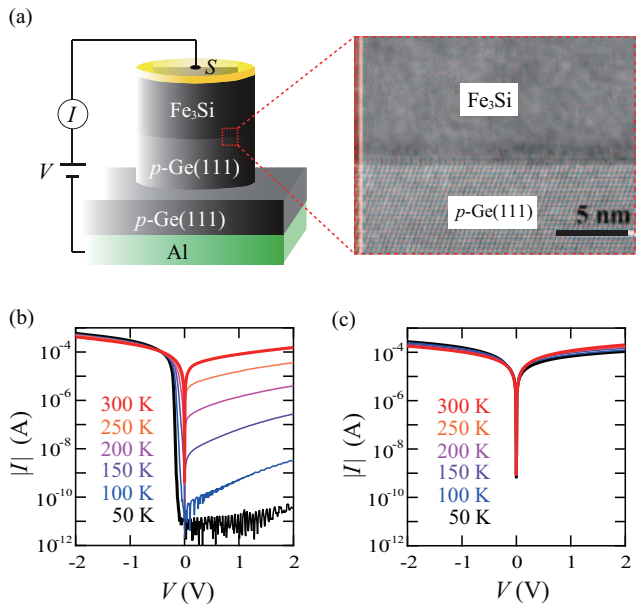}
\caption{(Color online) (a) The schematic illustration of the fabricated junctions with various areas ($S$). The enlarged picture is a cross-sectional transmission electron micrograph of the atomically controlled Fe$_{3}$Si/Ge(111) interface. (b), (c) $I$-$V$ characteristics of representative two junctions with $S =$ $\sim$1 $\mu$m$^{2}$. (b) Marked rectifying and (c) ohmic behavior with the temperature evolution.} 
\end{figure}  

In this Letter, we experimentally study the mechanism of the strong FLP at metal/Ge interfaces using the atomically matched Fe$_{3}$Si/Ge(111) heterostructure. For the high-quality Fe$_{3}$Si/$p$-Ge(111) junctions, we can easily observe the clear Schottky characteristics with a measurable Schottky barrier height, depending on the contact area of the Fe$_{3}$Si/Ge(111) junction. These transport data can be understood by the model based on the influence of the interfacial defects such as dangling bonds, and indicate that one should distinguish between intrinsic and extrinsic mechanisms for discussing the formation of the Schottky barrier at metal/Ge interfaces. This study will be developed for understanding FLP for almost all the metal/semiconductor interfaces. 
\begin{figure}[t]
\includegraphics[width=8.5cm]{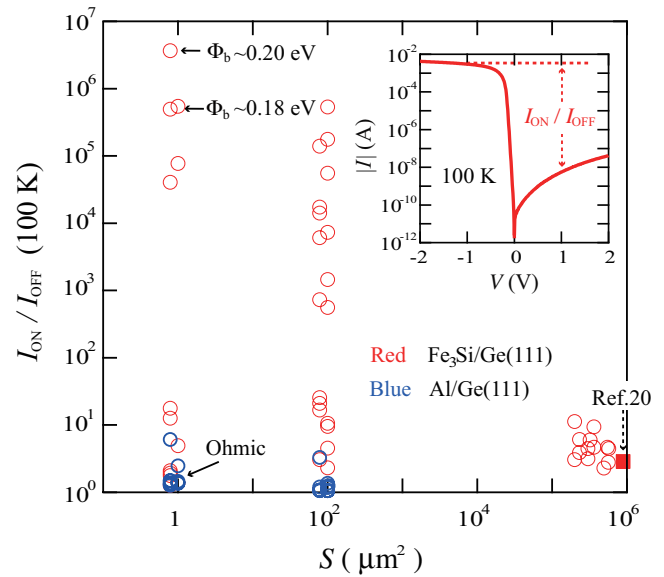}
\caption{(Color online) Summary of $I_\text{ON}$/$I_\text{OFF}$ ratio measured at 100 K for various Fe$_{3}$Si/$p$-Ge(111)/Al junctions (red) with $S =$ $\sim$1, $\sim$10$^{2}$, and $\sim$10$^{6}$ $\mu$m$^{2}$, together with that for Al/$p$-Ge(111)/Al (blue) junctions. The red-square symbol at $\sim$10$^{6}$ $\mu$m$^{2}$ is the data for a sample shown in Ref. \cite{Yamane}. The inset shows an $I$-$V$ characteristic for a junction with $S =$ $\sim$10$^{2}$ $\mu$m$^{2}$ at 100 K. The $I_\text{ON}$/$I_\text{OFF}$ ratio is defined at $V =$ $\pm$1 V. } 
\end{figure}

Prior to the fabrication of the Fe$_{3}$Si/Ge(111) heterostructure, $p$-Ge(111) substrates were chemically cleaned to remove contamination and native oxide from the surface, where the $p$-Ge(111) substrates were already on the market ($p$ = 9 $\times$ 10$^{14}$ cm$^{-3}$). The cleaned substrates were loaded  immediately into an ultra high vacuum chamber with a base pressure of $\sim$ 10$^{-7}$ Pa. After the surface heat treatment conducted at 550 $^{\circ}$C for 20 min, the substrate temperature was reduced down to 200$^{\circ}$C. An atomically clean surface was confirmed by observing the reflection high-energy electron diffraction (RHEED) patterns. Using low-temperature molecular beam epitaxy (MBE),\cite{Hamaya} we grew the epitaxial Fe$_{3}$Si layer with a thickness of 50 nm on $p$-Ge(111). During the growth, two-dimensional epitaxial growth was confirmed by the observation of RHEED patterns. After the growth, we fabricated Schottky diodes with a contact area $S$ ($S =$ $\sim$ 1, $\sim$10$^{2}$, and $\sim$10$^{6}$ $\mu$m$^{2}$) to examine electrical properties of the Fe$_{3}$Si/$p$-Ge(111) junctions.\cite{Yamane} Here all the diodes used in this study were fabricated from one Fe$_{3}$Si/$p$-Ge(111) wafer. The schematic diagram of the fabricated Fe$_{3}$Si/$p$-Ge/Al Schottky junctions is shown in Fig. 1(a), where a backside Al is an Ohmic contact. The Fe$_{3}$Si/Ge(111) interface is atomically flat,\cite{Yamane,Hamaya} as shown in a cross-sectional transmission electron micrograph (TEM) image in Fig. 1(a). To compare the atomically matched high-quality junctions with other mismatched ones, we also fabricated Al/$p$-Ge(111) junctions with the atomically mismatched interface by using the same MBE equipment after the same surface cleaning. The growth temperature of the Al/$p$-Ge(111) junctions was less than 25 $^{\circ}$C. 
  
First of all, we have studied $I$-$V$ characteristics for lots of Fe$_{3}$Si/$p$-Ge(111) junctions with $S =$ $\sim$1 $\mu$m$^{2}$. Figures 1(b) and 1(c) show representative $I$-$V$ characteristics at various temperatures. Despite the directly connected metal/Ge structure, we observe marked rectifying behavior in Fig. 1(b) at low temperatures surprisingly. The ratio of the forward current ($I_\text{ON}$) to the reverse current ($I_\text{OFF}$) reaches $\sim$ 10$^{8}$ at 50 K. The reverse bias current ($V >$ 0) increases with increasing temperature, implying the presence of an evident Schottky barrier. A rectifying $I$-$V$ characteristic can be obtained even at 300 K. Using the thermionic emission theory,\cite{Sze} we can roughly obtain $\Phi$$_{b}$ for hole transport. From the $\it{I}$-$\it{T}$ characteristics,\cite{Yamane} the $\Phi$$_{b}$ for holes in Fig. 1(b) was estimated to be $\sim$ 0.18 eV, which is the first experimental observation of $\Phi$$_{b}$ in the directly connected metal/$p$-Ge junctions. In contrast, almost ohmic characteristics are seen in Fig. 1(c) even at low temperatures, consistent with the strong FLP.\cite{Dimoulas,Toriumi} This means that the tunneling conduction occurs through the very low and thin Schottky barrier. When we decreased temperature from 50 to 10 K, we observed very small rectifications in $I$-$V$ characteristics. Note that despite almost the same devices fabricated from the same wafer there are different $I$-$V$ characteristics for Fe$_{3}$Si/$p$-Ge(111) junctions with $S =$ $\sim$1 $\mu$m$^{2}$. We also measured the $I$-$V$ characteristics of the Al/$p$-Ge(111) junctions with $S =$ $\sim$1 $\mu$m$^{2}$, measured at various temperatures, and then, almost all the junctions showed ohmic $I$-$V$ characteristics, also consistent with the strong FLP.\cite{Dimoulas,Toriumi} For the Al/$p$-Ge(111) junctions, we sometimes observed very small rectifications in $I$-$V$ characteristics as the temperature was decreased down to 10 K. However, we could not estimate $\Phi$$_{b}$ for Al/$p$-Ge(111). We emphasize that clear Schottky characteristics were achieved only when we used the atomically matched Fe$_{3}$Si/Ge(111) interface. 
\begin{figure}[t]
\includegraphics[width=7cm]{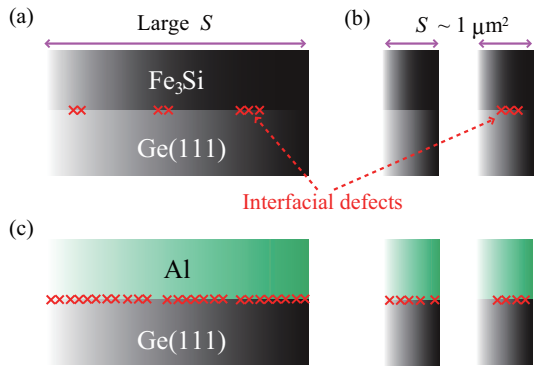}
\caption{(Color online) The schematic diagrams of the model for the Fe$_{3}$Si/Ge(111) interface with (a) large $S$ and (b) small $S$. There are some interfacial defects even at the atomically matched interface. (c) The diagrams in the case of the atomically mismatched Al/Ge(111) interface.} 
\end{figure}  

To understand the interesting features described above, we further examined the effect of $S$ on $I$-$V$ characteristics for many junctions in detail. As a consequence, we easily observed rectifying behavior in the $I$-$V$ curves at low temperatures for many Fe$_{3}$Si/$p$-Ge(111) junctions. In Fig. 2 we summarize $I_\text{ON}$/$I_\text{OFF}$ at 100 K for various Fe$_{3}$Si/$p$-Ge(111) junctions with $S =$ $\sim$1, $\sim$10$^{2}$, and $\sim$10$^{6}$ $\mu$m$^{2}$, together with the data for Al/$p$-Ge(111) junctions. Here for all kinds of $S$ we measured $I$-$V$ curves of more than fifteen junctions, and the $I_\text{ON}$/$I_\text{OFF}$ values were roughly defined as the data at $V =$ $\pm$1 V, as shown in the inset. In the main figure we can reconfirm that, for the Al/$p$-Ge(111) junctions with the atomically mismatched interfaces, all the $I_\text{ON}$/$I_\text{OFF}$ values are less than 10$^{1}$ even for the smallest junctions with $S =$ $\sim$1 $\mu$m$^{2}$, consistent with the strong FLP. On the other hand, for the Fe$_{3}$Si/$p$-Ge(111) junctions, we can evidently see the difference in the $I_\text{ON}$/$I_\text{OFF}$ values depending strongly on $S$. For the large junctions ($S =$ $\sim$10$^{6}$ $\mu$m$^{2}$), the $I_\text{ON}$/$I_\text{OFF}$ values are almost constant ($\sim$ 10$^{1}$). The small rectification of the $I$-$V$ curves disappeared at $\sim$170 K, similar to our previous work.\cite{Yamane} In contrast, for the small junctions with $S =$$\sim$1 and $\sim$10$^{2}$ $\mu$m$^{2}$, the $I_\text{ON}$/$I_\text{OFF}$ values are scattered from the wide range 10$^{0}$ $\le$ $I_\text{ON}$/$I_\text{OFF}$ $\le$ 10$^{7}$. For $S =$ $\sim$1 $\mu$m$^{2}$ the junctions shown in Figs. 1(b) and 1(c), i.e., $\Phi$$_{b}$ $\sim$ 0.18 eV and ohmic, respectively, are also plotted in Fig. 2 (see arrows), and the junction with $\Phi$$_{b}$ $\sim$ 0.20 eV shows the largest $I_\text{ON}$/$I_\text{OFF}$ $\sim$ 7 $\times$ 10$^{6}$. The scattering of the $I_\text{ON}$/$I_\text{OFF}$ values in Fig. 2 implies that many kinds of the Schottky barriers are distributed for our Fe$_{3}$Si/$p$-Ge(111) junctions. We discuss it later. Note that the scattering range for $S =$ $\sim$1 $\mu$m$^{2}$ is narrow ($I_\text{ON}$/$I_\text{OFF}$ $<$ $\sim$10$^{1}$, $\sim$10$^{5}$ $<$ $I_\text{ON}$/$I_\text{OFF}$ $<$ 10$^{7}$) compared to $S =$ $\sim$10$^{2}$ $\mu$m$^{2}$ ($I_\text{ON}$/$I_\text{OFF}$ $<$ $\sim$10$^{2}$, 10$^{3}$ $<$ $I_\text{ON}$/$I_\text{OFF}$ $<$ 10$^{6}$). 
\begin{figure}[t]
\includegraphics[width=8.5cm]{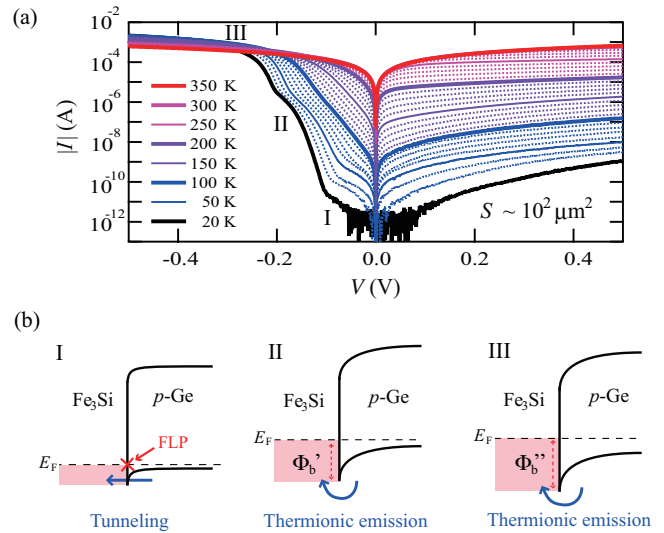}
\caption{(Color online) (a) Temperature evolution of $I$-$V$ characteristics with three-step staircase in forward bias ($V <$ 0), observed in Fe$_{3}$Si/Ge(111) junctions with $S =$ $\sim$10$^{2}$ $\mu$m$^{2}$. (b) Schematic illustrations of the the hole transport in $V <$ 0 at low temperatures. There are plural current channels in the Fe$_{3}$Si/$p$-Ge(111) junctions and all the channels contribute to the forward-bias current.} 
\end{figure}  

Hereafter we intensively discuss  the effect of $S$ on $I_\text{ON}$/$I_\text{OFF}$, presented in Fig. 2. For almost all the Al/$p$-Ge(111) junctions, we can regard the small $I_\text{ON}$/$I_\text{OFF}$ values of $\sim$ 10$^{0}$ at low temperatures as consequences of the strong influence of FLP on the hole transport. In short, there is almost no effect of $S$ on $I_\text{ON}$/$I_\text{OFF}$ at low temperatures. On the other hand, there is the marked effect of $S$ on $I_\text{ON}$/$I_\text{OFF}$ for the Fe$_{3}$Si/$p$-Ge(111) junctions. Even if we use the atomically matched Fe$_{3}$Si(111)/Ge(111) junctions shown in Fig. 1 of Ref. \cite{Yamane}, there will be still some atomic imperfection at the (111) planes between Fe or Si in Fe$_{3}$Si and Ge, giving rise to the defects such as dangling bonds in Ge at the interface. Therefore, the presence of some unintentional defects at the Fe$_{3}$Si/$p$-Ge(111) interface can be considered generally. Assuming Poisson distribution for the defects at the Fe$_{3}$Si/Ge(111) interface, we can show a possible scenario considering the correlation between $S$ and interfacial defects as follows. When $S$ is large such as $S =$ $\sim$10$^{6}$ $\mu$m$^{2}$, the Fe$_{3}$Si/Ge(111) junction includes the some interfacial defects, as schematically illustrated in Fig. 3(a). The observed small $I_\text{ON}$/$I_\text{OFF}$ values of $\sim$ 10$^{1}$ in Fig. 2 should be regarded as consequences of the influence of the interfacial defects on hole transport properties. On the other hand, when we reduce $S$ as small as possible, we can encounter the two different interfaces, as shown in Fig. 3(b). Then, we can obtain almost no or a strong contribution of the defects to the transport properties. Surely, for $S =$ $\sim$1 $\mu$m$^{2}$, we have already obtained such two different contributions in Figs. 1(b) and (c). Also, we can regard the two discrete regions in Fig. 2, i.e., $I_\text{ON}$/$I_\text{OFF}$ $<$ $\sim$10$^{1}$ and $\sim$10$^{5}$ $<$ $I_\text{ON}$/$I_\text{OFF}$ $<$ 10$^{7}$, as the above two different contributions. In contrast, since there are a large number of  interfacial detects at the atomically mismatched Al/$p$-Ge(111) interfaces, as illustrated in Fig. 3(c), we cannot encounter the situation with almost no interfacial defect even if we use the small junctions which we can fabricate. Since we have not obtained the data with Schottky characteristics even for the smallest Al/$p$-Ge(111) junctions, the above scenario based on the influence of the interfacial defects on the hole transport can also be applied to the atomically mismatched interfaces. 

Recent papers also indicated that passivation techniques of the Ge surface can work effectively for changing electrical properties of the metal/Ge junctions.\cite{Thathachary,Lieten2,Zhou2} Also, theoretical studies suggested that the defect levels lie just above the valence band,\cite{Weber,Broqvist} indicating that the previous experimental results shown in Ref. \cite{Dimoulas,Toriumi} originate from the interfacial defects such as dangling bonds. These studies strongly support our new findings in this paper. We want to emphasize that lots of researchers have so far observed the predominant influence of interfacial defects on the transport properties of metal/Ge junctions. In other words, this study indicates that if one discusses the fundamental mechanism of the formation of the Schottky barrier at metal/Ge interfaces, one should consider the influence of interfacial defects, i.e., extrinsic mechanisms. In the future, we will need further exploration of the method for reducing the number of the defects formed at the atomically matched metal/Ge interfaces.

Finally, we discuss the scattering of the $I_\text{ON}$/$I_\text{OFF}$ values in Fig. 2 for $S =$ $\sim$1 and $\sim$10$^{2}$ $\mu$m$^{2}$. For $S =$ $\sim$1 $\mu$m$^{2}$, the various Fe$_{3}$Si/$p$-Ge(111) interfaces with Schottky barriers from $\Phi$$_{b}$ $\sim$ 0.20 eV to almost ohmic were demonstrated. Also, the scattering range for $S =$ $\sim$1 $\mu$m$^{2}$ is narrower than that for $S =$ $\sim$10$^{2}$ $\mu$m$^{2}$. This feature indicates that the scattering of the $I_\text{ON}$/$I_\text{OFF}$ values is related to $S$, i.e., the number of interfacial defects. Since the $I_\text{ON}$/$I_\text{OFF}$ value is usually attributed to $\Phi$$_{b}$, we should interpret that $\Phi$$_{b}$ is also related to the number of defects. Thus, somewhat different contributions of the number of defects to $\Phi$$_{b}$ can be considered in the Fe$_{3}$Si/Ge(111) junctions. On the basis of this consideration, we show new findings for electrical properties in the Fe$_{3}$Si/Ge(111) junctions. Figure 4(a) displays the $I$-$V$ curves for an Fe$_{3}$Si/$p$-Ge(111) junction with $S =$ $\sim$10$^{2}$ $\mu$m$^{2}$ at various temperatures. Interestingly, we can see the staircase-like $I$-$V$ characteristics in forward bias ($V <$ 0), denoted by I, II, and III. In this study, we frequently observed these staircase-like $I$-$V$ characteristics for $S =$ $\sim$10$^{2}$ $\mu$m$^{2}$ at low temperatures though the room-temperature $I$-$V$ curves were almost the same ohmic characteristics. The interpretation of the staircase-like $I$-$V$ characteristics is shown in Fig. 4(b). That is, there are plural current channels for holes, resulting from the tunneling conduction due to FLP (I) and from the thermionic-emission conductions (II, III) with two different Schottky barriers with $\Phi$$_{b}$$^{'}$ and $\Phi$$_{b}$$^{''}$. At low temperatures the forward bias current ($V <$ 0) can detect the contributions of all the current channels, while, with increasing temperature, $\Phi$$_{b}$$^{'}$ and $\Phi$$_{b}$$^{''}$ cannot affect the transport properties, giving rise to the disappearance of the staircase-like behavior. This feature means that there are different contributions of the interfacial defects to $\Phi$$_{b}$ for just one Fe$_{3}$Si/$p$-Ge(111) interface in the middle-sized junctions. By using the atomically matched Fe$_{3}$Si/$p$-Ge(111) interface, we can evidently observe some new findings for electrical properties of metal/Ge interfaces. This study indicates that one should distinguish between intrinsic and extrinsic mechanisms for discussing the formation of the Schottky barrier at metal/Ge interfaces.

In summary, we have studied the mechanism of FLP at metal/Ge junctions using the atomically matched Fe$_{3}$Si/Ge(111) interface.  Despite metal/$p$-Ge interfaces, we clearly observed rectifying current-voltage characteristics with a measurable Schottky barrier height, depending on the contact area of the Fe$_{3}$Si/Ge(111) junction. These data can be understood by the model based on the influence of the interfacial defects such as dangling bonds, and indicate that one should distinguish between intrinsic and extrinsic mechanisms for discussing the formation of the Schottky barrier at metal/Ge interfaces. This study will give us an important principle for understanding FLP for almost all the metal/semiconductor interfaces, which has been studied over a long period. 

K.H. and M.M. acknowledge Prof. H. Nakashima for useful discussions. K.K. and S.Y. acknowledge JSPS Research Fellowships for Young Scientists. This work was partly supported by Industrial Technology Research Grant Program from NEDO and Grant-in-Aid for Young Scientists (A) from JSPS. 



\begin{thebibliography}{3}
\bibitem{Heine}
V. Heine, Phys. Rev. {\bf 138}, A1689 (1965).
\bibitem{Rowe}
J. E. Rowe, S. B. Christman, and G. Margaritondo, Phys. Rev. Lett. {\bf 35}, 1471 (1975).
\bibitem{Spicer}
W. E. Spicer, I. Lindau, P. Skeath, and C.Y. Su, J. Vac. Sci. Technol. {\bf 17}, 1019 (1980).
\bibitem{Tung}
R. T. Tung, Phys. Rev. Lett. {\bf 52}, 461 (1984).
\bibitem{Hasegawa}
H. Hasegawa and H. Ohno, J. Vac. Sci. Technol. B {\bf 4}, 1130 (1986).
\bibitem{Tersoff}
J. Tersoff, Phys. Rev. Lett. {\bf 52}, 465 (1984); Phys. Rev. B {\bf 32}, 6968 (1985).
\bibitem{Heslinga}
D. R. Heslinga, H. H. Weitering, D. P. van der Werf, T. M. Klapwijk, and T. Hibma, Phys. Rev. Lett. {\bf 64}, 1589 (1990). 
\bibitem{Tung2}
R. T. Tung, Phys. Rev. Lett. {\bf 84}, 6078 (2000); Phys. Rev. B {\bf 64}, 205310 (2004).
\bibitem{Connelly}
D. Connelly, C. Faulkner, P.A. Clifton, and D. E. Groupp, Appl. Phys. Lett. {\bf 88}, 012105 (2006).
\bibitem{Mohney}
A. Dimoulas, A. Toriumi, and S. E. Mohney, MRS bulletin {\bf 34}, 522 (2009); J. Hu, H.-S. P. Wong, K. Saraswat, MRS bulletin {\bf 36}, 112 (2011).
\bibitem{Lee}
T. Tezuka, S. Nakaharai, Y. Moriyama, N. Sugiyama, and S. Takagi, IEEE Electron Device Lett. {\bf 26}, 243 (2005); M. Miyao, K. Toko, T. Tanaka, and T. Sadoh, Appl. Phys. Lett. {\bf 95}, 022115 (2009); C. H. Lee, T. Nishimura, T. Tabata, S. K. Wang, K. Nagashio, K. Kita, and A. Toriumi, IEDM Tech. Dig., 416 (2010).
\bibitem{Dimoulas}
A. Dimoulas, P. Tsipas, A. Sotiropoulos, and E. K. Evangelou, Appl. Phys. Lett. {\bf 89}, 252110 (2006).
\bibitem{Toriumi}
T. Nishimura, K. Kita, and A. Toriumi, Appl. Phys. Lett. {\bf 91}, 123123 (2007).
\bibitem{Nishimura}
T. Nishimura, K. Kita, and A. Toriumi, Appl. Phys. Exp. {\bf 1}, 051406 (2008).
\bibitem{Kobayashi}
M. Kobayashi, A. Kinoshita, K. Saraswat, H.-S. Philip Wong, and Y. Nishi, J. Appl. Phys. {\bf 105}, 023702 (2009).
\bibitem{Wager}
J. F. Wager and J. Robertson, J. Appl. Phys. {\bf 109}, 094501 (2011).
\bibitem{Lieten}
R. R. Lieten, S. Degroote, M. Kuijk, and G. Borghs, Appl. Phys. Lett. {\bf 92}, 022106 (2008).
\bibitem{Zhou2}
Y. Zhou, W. Han, Y. Wang, F. Xiu, J. Zou, R. K. Kawakami, and K. L. Wang, Appl. Phys. Lett. {\bf 96}, 102103 (2010).
\bibitem{Thathachary}
A. V. Thathachary, K. N. Bhat, N. Bhat, and M. S. Hegde, Appl. Phys. Lett. {\bf 96}, 152108 (2010).
\bibitem{Yamane}
K. Yamane, K. Hamaya, Y. Ando, Y. Enomoto, K. Yamamoto, T. Sadoh, and M. Miyao, Appl. Phys. Lett. {\bf 96}, 162104 (2010).
\bibitem{Lieten2}
R. R. Lieten, V. V. Afanas'ev, N. H. Thoan, S. Degroote, W. Walukiewicz, and G. Borghs, J. Electrochem. Soc. {\bf 158}, H358 (2011).
\bibitem{Moss}
J. Moss and P. J. Brown, J. Phys. F: Metal Phys. {\bf 2}, 358 (1972).
\bibitem{Niculescu}
V. A. Niculescu, T. J. Burch, and J. I. Budnick, J. Magn. Magn. Mater. {\bf 39}, 223 (1983).
\bibitem{Hamaya}
K. Hamaya, Y. Ando, T. Sadoh, and M. Miyao, Jpn. J. Appl. Phys. {\bf 50}, 010101 (2011); K. Hamaya, T. Murakami, S. Yamada, K. Mibu, and M. Miyao, Phys. Rev. B {\bf 83}, 144411 (2011).
\bibitem{Sze}
S. M. Sze, Physics of Semiconductor Devices, 2nd ed. (Wiley, New York 1981), pp. 270-286. 
\bibitem{Weber}
J. R. Weber, A. Janotti, P. Rinke, and C. G. Van de Walle, Appl. Phys. Lett. {\bf 91}, 142101 (2007).
\bibitem{Broqvist}
P. Broqvist, A. Alkauskas, and A. Pasquarello, Phys. Rev. B {\bf 78}, 075203 (2008).



\end{thebibliography}
\end{document}